\documentclass[9pt,twocolumn,twoside,nostamp,dvipsnames]{pnas-new}
\templatetype{pnasresearcharticle}

\usepackage[utf8]{inputenc}
\usepackage[T1]{fontenc}
\usepackage{microtype}
\usepackage{graphicx}
\usepackage{amsmath}
\usepackage{bm}
\usepackage{subcaption}
\usepackage{aas_macros}

\def\hmpc{h^{-1}{\rm Mpc}}
\def\hgpc{\;h^{-1}{\rm Gpc}}

\def\hkpc{h^{-1}\, {\rm kpc}}

\def\msun{\, M_{\odot}}

\def\simlt{\lower.5ex\hbox{$\; \buildrel < \over \sim \;$}}
\def\simgt{\lower.5ex\hbox{$\; \buildrel > \over \sim \;$}}

\makeatletter

\newcommand{\Rmnum}[1]{\expandafter\@slowromancap\romannumeral #1@}
\makeatother

\renewcommand{\d}{\mathrm{d}}

\title{AI-assisted super-resolution cosmological simulations}

\author[a,1]{Yin Li}
\author[b,c,1]{Yueying Ni}
\author[b,c]{Rupert A.~C. Croft}
\author[b,c]{Tiziana Di Matteo}
\author[d]{Simeon Bird}
\author[e]{Yu Feng}

\affil[a]{Center for Computational Astrophysics \& Center for Computational Mathematics,
Flatiron Institute, 162 5th Avenue, New York, NY 10010}
\affil[b]{McWilliams Center for Cosmology, Department of Physics,
Carnegie Mellon University, Pittsburgh, PA 15213}
\affil[c]{NSF AI Planning Institute for Physics of the Future,
Carnegie   Mellon  University, Pittsburgh, PA 15213, USA}
\affil[d]{Department of Physics and Astronomy, University of California Riverside,
900 University Ave, Riverside, CA 92521}
\affil[e]{Berkeley Center for Cosmological Physics and Department of Physics,
University of California, Berkeley, CA 94720}

\leadauthor{Li \& Ni}

\correspondingauthor{\textsuperscript{1}
E-mail: eelregit@gmail.com \& yueyingn@andrew.cmu.edu}

\authorcontributions{Author contributions:
YL, YN, RACC, TDM, SB, and YF designed research;
YL and YN performed research;
YL, YN, SB, and YF contributed software;
and YL, YN, RACC, TDM, SB, and YF wrote the paper.
}

\authordeclaration{The authors declare no conflict of interest.}

\dates{This manuscript was compiled on \today}

\doi{\url{www.pnas.org/cgi/doi/10.1073/pnas.XXXXXXXXXX}}


\significancestatement{
Cosmological simulations are indispensable for understanding
our Universe, from the creation of cosmic web
to the formation of galaxies and their central black holes.
This vast dynamic range incurs large computational costs,
demanding sacrifice of either resolution or size, and often both.
We build a deep neural network to enhance low-resolution dark matter simulations,
generating super-resolution realizations that agree remarkably well with authentic high-resolution counterparts
on their statistical properties, and are orders-of-magnitude faster.
It readily applies to larger volumes, and generalizes to rare
objects not present in the training data.
Our study shows that deep learning and cosmological simulations can be a powerful combination,
to model the structure formation of our Universe over its full dynamic range.
}

\begin{abstract}
Cosmological simulations of galaxy formation are limited by finite
computational resources.
We draw from the ongoing rapid advances in Artificial Intelligence
(specifically Deep Learning) to address this problem.
Neural networks have been developed to learn from high-resolution (HR)
image data, and then make accurate super-resolution (SR) versions of
different low-resolution (LR) images.
We apply such techniques to LR cosmological N-body simulations, generating
SR versions.
Specifically, we are able to enhance the simulation resolution by
generating 512 times more particles and predicting their displacements
from the initial positions.
Therefore our results can be viewed as new simulation realizations
themselves rather than projections, e.g., to their density fields.
Furthermore, the generation process is stochastic, enabling us to sample
the small-scale modes conditioning on the large-scale environment.
Our model learns from only 16 pairs of small-volume LR-HR simulations, and is then
able to generate SR simulations that successfully reproduce
the HR matter power spectrum to percent level up to $16\,\hmpc$,
and the HR halo mass function to within $10 \%$ down to $10^{11} \msun$.
We successfully deploy the model in a box 1000 times larger than the
training simulation box, showing that high-resolution mock surveys
can be generated rapidly. We conclude that AI assistance
has the potential to revolutionize modeling of small-scale galaxy
formation physics in large cosmological volumes.
\end{abstract}

\begin{document}

\maketitle
\thispagestyle{firststyle}
\ifthenelse{\boolean{shortarticle}}{\ifthenelse{\boolean{singlecolumn}}{\abscontentformatted}{\abscontent}}{}

\dropcap{A}s telescopes and satellites become more powerful, observational data on
galaxies, quasars and the matter in intergalactic space becomes more detailed,
and covers a greater range of epochs and environments in the Universe.
Our cosmological simulations \citep[see e.g.,][]{vogelsberger20} must also become more detailed and more wide
ranging in order to make predictions and test the effects of different physical
processes and different dark matter candidates.
Even with supercomputers we are forced to decide whether to maximize either
resolution, or volume, or else compromise on both.
These limitations can be overcome through the development of methods
that leverage techniques from the Artificial Intelligence (AI) revolution
\citep[see e.g.,][]{russell2020}, and make super-resolution (SR) simulations possible.
In the present work we begin to explore this possibility, combining knowledge and existing super-scalable codes for petascale-plus
cosmological simulations \cite{Feng2015} with Machine Learning (ML) techniques to
effectively create representative volumes of the Universe that
incorporate information from higher-resolution models of galaxy formation.
Our first attempts, presented here, involve simulations with dark matter
and gravity only, and extensions to full hydrodynamics will follow.
This hybrid approach which will imply offloading simulations to Neural
Networks (NN) and other ML algorithms has the promise to enable the prediction of quasar,
supermassive black hole and galaxy properties in a way which is statistically
identical to full hydrodynamic models but with a significant speed up.

Adding details to images below the resolution scale (SR image
enhancement) has become possible with the latest advances in Deep Learning \citep[DL,
ML with NN;][]{goodfellow2016deep}, including Generative Adversarial
Networks \citep[GANs;][]{goodfellow2014generative}.
The technique has applications in many fields, from microscopy to law
enforcement \citep{wang2020}.
It has been used for observational astronomical images by \cite{schawinski},
to recover galaxy features from below the resolution scale in degraded HST
images.
Besides SR image enhancement, DL has started to find applications in
cosmological simulations.
For example, \cite{he2019learning} and \cite{berger2019} showed how NNs can predict the nonlinear formation of structures given simple linear theory predictions.
NN models have also been trained to predict galaxies \citep{Modi2018, zhang2019} and 21 cm emission from neutral hydrogen \citep{wadekar2020hinet},
from simulations that only contain dark matter.
GANs have been used in \citep{2018ComAC...5....4R} to generate image slices of cosmological models, and to generate dark matter halos from density fields \citep{ramanah2019painting}.
ML techniques other than DL find many applications too. For example,
\cite{kamdar2016machine} have applied extremely randomized trees to dark matter simulations to predict hydrodynamic galaxy properties.

\begin{figure*}
\centering
  \includegraphics[width=1\textwidth]{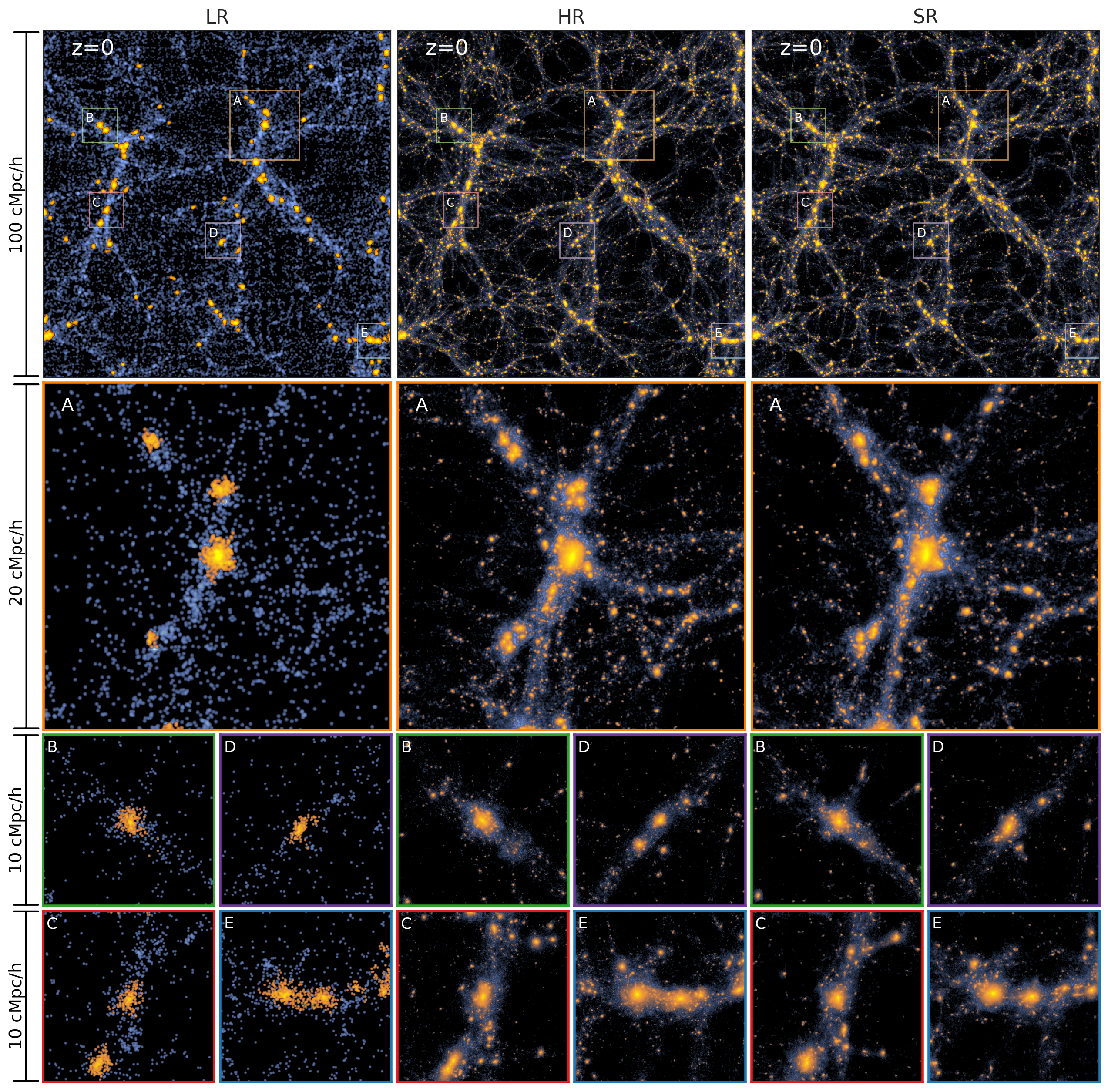}
  \caption{2D projections of the low-, high-, and super-resolution (LR, HR, and SR)
    dark matter density fields at $z=0$.
    The blue background shows the smoothed density field of all the dark
    matter particles. The particles in FOF groups are highlighted in orange
    to help visually identify the halos.
    The top panels show slabs from the full box of 100 $\hmpc$ side length
    and 20 $\hmpc$ thickness.
    The middle panels zoom into the orange boxes (A) from the top panel,
    each of size $(20\,\hmpc)^3$.
    The bottom two rows show the four zoom-in boxes (B, C, D, and E) which are $(10\,\hmpc)^3$ in size to reveal even finer details.
    The first two columns show the LR and HR simulations, which have the same initial conditions but a factor of $512\times$ different mass resolution.
    The rightmost column shows \emph{one} of the SR realizations generated
    by our trained model.
    All density projections are smoothed by a Gaussian filter on a scale of 5 $\hkpc$,
    using \texttt{gaepsi2}
    (\url{https://github.com/rainwoodman/gaepsi2}).
  }\label{fig:slices}
\end{figure*}

Generating mocks for future sky surveys requires large volumes and high accuracy, a task that quickly becomes computationally prohibitive.
To alleviate the cost, recently \cite{Dai2020b} developed a Lagrangian based parametric ML model to predict various hydrodynamical outputs from the dark matter density field.
In other work, \cite{Dai2018,Dai2020a} sharpened the particle distribution using a potential gradient descent (PGD) method starting from low-resolution simulations.
Note however these approaches did not aim to enhance the spatial or mass resolution of a simulation.

On the DL side, recently \cite{ramanah20} has explored using the SR technique to map
density fields of LR cosmological simulations to that of the HR ones.
While the goal is similar, our work has the following three key
differences.
First, instead of focusing on the dark matter density field, we aim to
enhance the number of particles and predict their displacements, from
which the density fields can be inferred.
This approach allows us to preserve the particle nature of the N-body
simulations, therefore to interpret the SR outputs as simulations themselves.
Second, we test our technique at a higher SR ratio.
Compared to \cite{ramanah20} that increased the number of Eulerian voxels
by 8 times, we increase the number of particles and thus the mass
resolution by a factor of 512.
Finally, to facilitate future applications of SR on hydrodynamic
simulations in representative volumes, we test our method at much smaller
scales, and in large simulations whose volume is much bigger
than that of the training data.

\section*{Results}\label{sec:results}

In this work we employ DL, the class of machine learning
algorithms involving artificial neural networks, to progressively
transform a LR input simulation via multiple layers of
neurons into an output that statistically reproduces the HR target.
We build and train a DL model based on physical considerations,
and use the displacements of particles in the N-body
simulations as inputs and outputs.
This allows us to interpret the results as simulation realizations themselves.
To help train the generative network, we employ a GAN approach to
simultaneously train a discriminative network that contests with the
generator in a game by evaluating its outputs.
We validate our model by comparing the generated SR
simulations to the authentic HR simulations both visually and quantitatively,
using summary statistics including the power spectrum and the halo mass function.

\subsection*{Visual comparison}\label{sub:visual}

\begin{figure}[t]
  \centering
  \includegraphics[width=1\columnwidth]{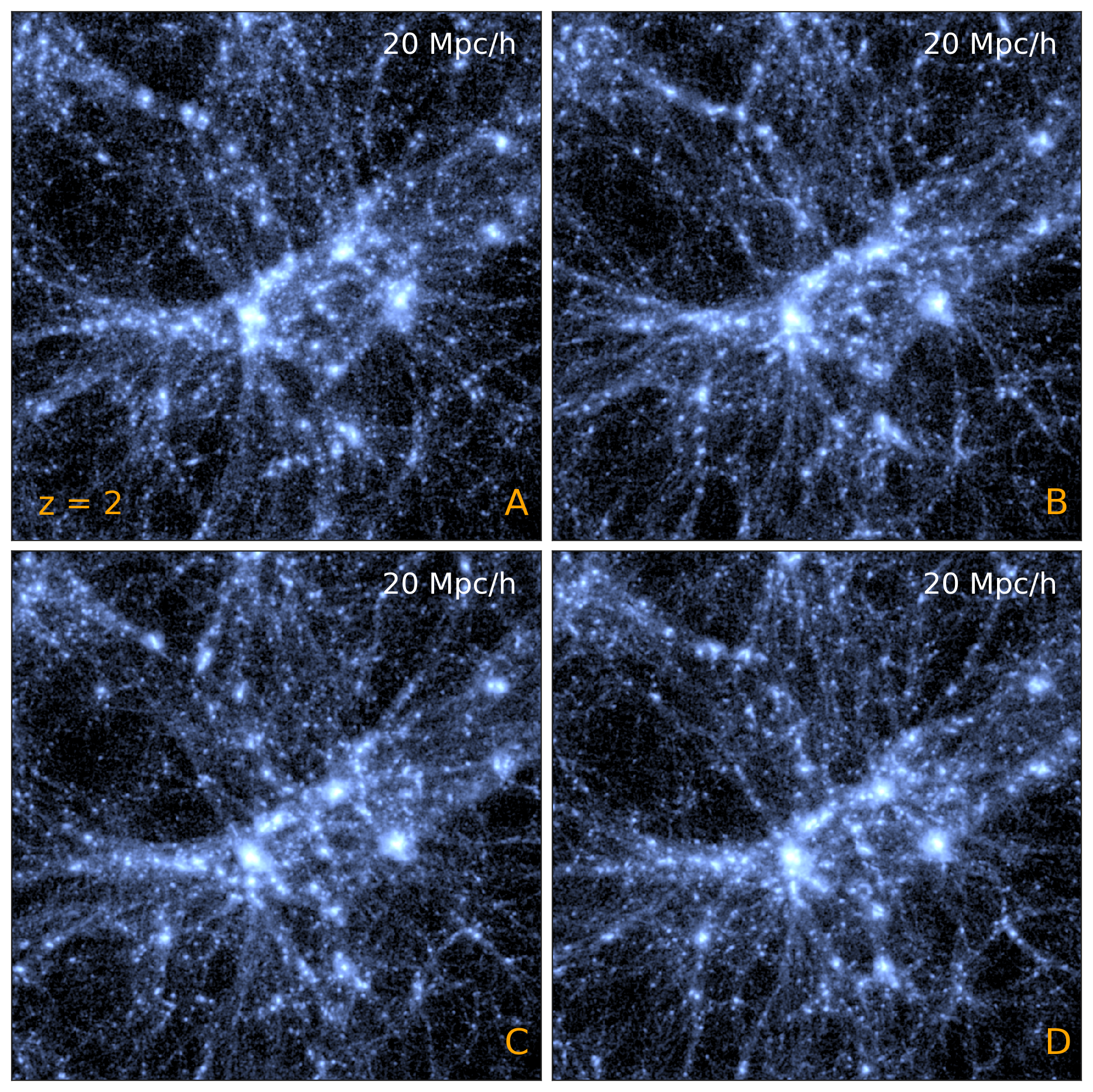}
  \caption{Using our GAN-based algorithm, we can generate different SR realizations from the same LR inputs.
  We show one HR and three random SR projections,
  of the same $(20\,\hmpc)^3$ regions and the same LR field,
  at $z=2$.
  To demonstrate their remarkable similarities, we intentionally omit the HR and SR labels,
  and invite the readers to guess \emph{which panel is HR}
  before checking the answer in the footnote\protect\footnotemark.
  The HR and three SR realizations have the same
  large-scale structures but are all different in their small-scale features.
  Yet they appear statistically indistinguishable.
  The color scheme and smoothing method are the same as in Fig.~\ref{fig:slices}.
  }\label{fig:Realizations}
\end{figure}

First let us visually compare the generated SR results to the authentic
HR simulations.
We obtain the SR particle positions by moving them from their original
grid locations using the predicted displacements.
In the three columns of Fig.~\ref{fig:slices}, respectively, we show the LR, HR, and SR dark matter density fields at $z=0$.
We visualize all dark matter particles in blue and highlight the
FOF halos (see below) in orange.

As shown in Fig.~\ref{fig:slices}, our SR technique is able to
recover sharp features and small-scale structures starting from inputs
of extremely poor resolution.
It forms halos where the input LR can not resolve them, e.g., in the figure
the LR can only form the most massive handful of halos in the box whereas
SR is able to improve the halo mass range by orders of magnitude.
The SR outputs trace the large-scale environment set by the LR inputs,
while appearing statistically almost indistinguishable from the
HR targets on the small scales.

Our method also allows us to map the same LR input to multiple outputs.
We achieve this by adding noise between layers in the generator as
explained in the Materials and Methods.
In Fig.~\ref{fig:Realizations}, we show one HR and three random SR projections,
of the same $(20\,\hmpc)^3$ regions and the same LR field,
at $z=2$.
As expected, above and around the input resolution scales
the HR and SR fields are consistent with each other,
while being apparently different but visually similar on smaller scales.
To demonstrate their remarkable similarities, we
invite the readers to guess which panel is HR
before checking the answer in the footnote\footnotemark[\value{footnote}].

\footnotetext{The answer to Fig.~\ref{fig:Realizations} is A.}

The visual tests have demonstrated that our SR technique
is able to not only generate small-scale structures that statistically resemble
the target, but also sample them conditioning
on their larger-scale environment.
These capabilities make our method extremely promising for a variety of
other SR applications.

\subsection*{Power spectrum}

\begin{SCfigure*}[\sidecaptionrelwidth][t]
  \centering
  \includegraphics[width=0.8\textwidth]{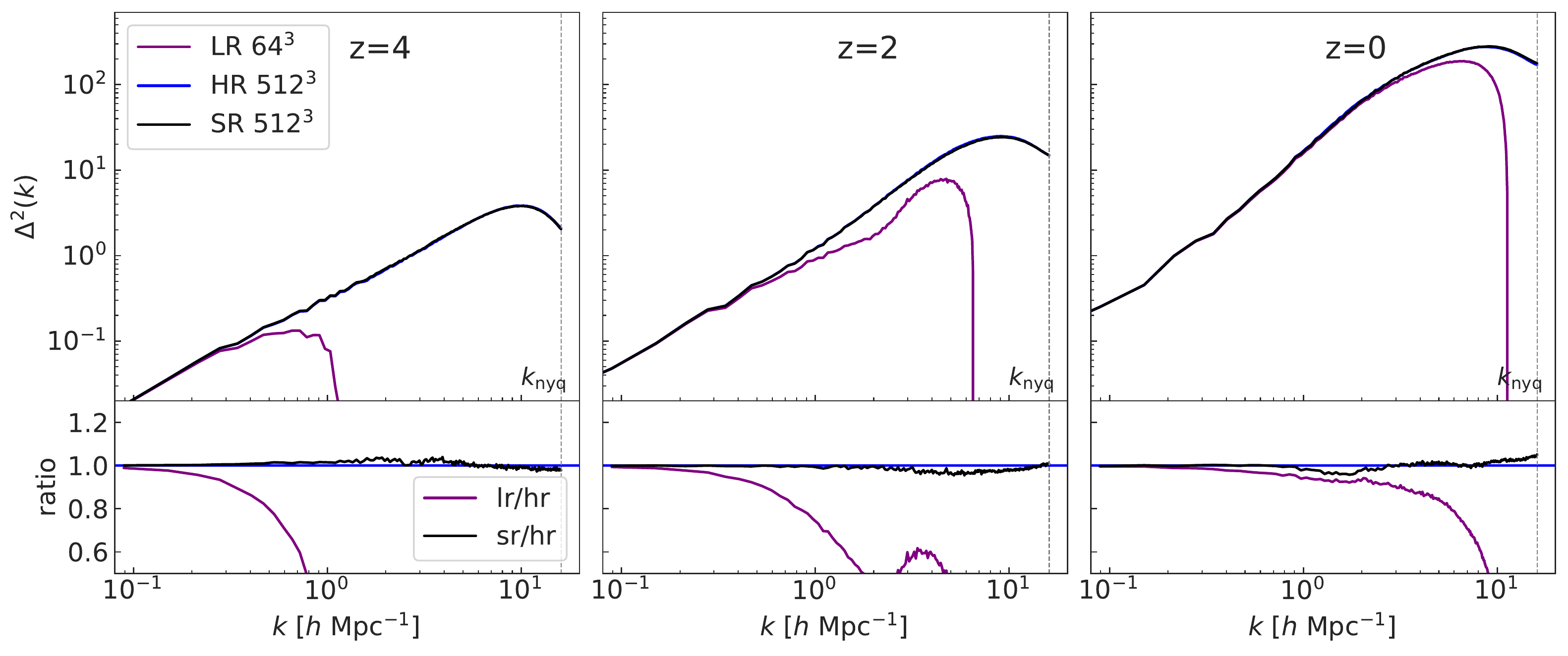}
  \caption{
  Dimensionless matter power spectrum $\Delta^2$ comparison
  on LR (purple), HR (blue), and SR (black) test realizations,
  at $z=4, 2,$ and $0$.
  The vertical dashed lines mark the Nyquist wavenumber.
  LR power quickly vanishes on small scales at $z=4$.
  This deficit persists and is partly compensated by formation of poorly
  resolved halos at $z=2$ and $z=0$.
  The SR result is a dramatic improvement, with the SR power spectra matching the HR curves remarkably well,
  within a few percent on all scales at all redshifts.
  }\label{fig:power}
\end{SCfigure*}

The matter power spectrum, that describes the amplitude of density
fluctuations as a function of scale, is perhaps the most commonly used summary
statistic in cosmology.
It describes in Fourier space the 2-point correlation, which completely
specifies the statistical properties of Gaussian random fields.
And to the best of our knowledge, the cosmological density field is well represented after the Big Bang by a Gaussian random field,
and remains so in the linear growth regime.
Therefore it is standard to compare the SR and HR simulations on their
matter power spectra $P(k)$.

For a periodic simulation box, the matter power spectrum can be measured
by a discrete Fourier transform
\begin{equation}
  P(k_i) = \frac1{N_i} \sum_{k_i < k \leq k_{i+1}}
    \frac{\bigl| \delta(\bm{k}) \bigr|^2} V,
\end{equation}
where $V$ is the simulation volume and $\delta(\bm{k})$ is the
discrete Fourier transform of the overdensity field $\delta(\bm{x})
\equiv \rho(\bm{x}) / \bar\rho - 1$, with $\rho(\bm{x})$ and $\bar\rho$
being the matter density field and its mean value, respectively.
Due to statistical isotropy, the power spectrum is only a function of
the wavenumber $k = |\bm{k}|$ and independent of the direction of $\bm{k}$.
The above estimator has already exploited this by averaging spherically
within each $k$-bin.
$N_i$ is the number of modes falling in the $i$-th bin $(k_i, k_{i+1}]$.

We compute the density field by assigning the particle mass to a
$512^3$ mesh using the CIC (Cloud-in-Cell) scheme, for all LR, HR, and SR simulations.
A common practice in power spectrum estimation is deconvolution of
the resampling window (here CIC) after the discrete Fourier transform \cite{Jing05}.
However, this amplifies the noises and leads to artifacts in the LR results.
Instead, using a large FFT grid for both resolutions, we avoid deconvolving the
resampling window and can compare power spectra from
different resolutions on an equal footing.

In Fig.~\ref{fig:power}, we compare the dimensionless power spectra
$\Delta^2(k) \equiv k^3 P(k) / 2 \pi^2$,
at $z=4, 2,$ and $0$.
The vertical dashed lines mark the Nyquist wavenumber $k_{\mathrm{nyq}} = \pi N_{\mathrm{mesh}}/L_{\mathrm{box}}$ with $N_{\mathrm{mesh}}=512$ and $L_{\mathrm{box}} = 100\,\hmpc$.
The monotonically increasing $\Delta^2(k)$ is a good proxy for the variance of
matter density as a function of scale, and thus a useful indicator that divides
the linear and nonlinear (towards increasing $k$) regimes,
by $k_\mathrm{NL}$ where $\Delta^2(k_\mathrm{NL}) = 1$.

Due to limited mass resolution, LR power quickly vanishes on small scales
at $z=4$.
This deficit persists and is partly compensated by formation of poorly
resolved halos at $z=2$ and $z=0$.
On the other hand, the SR power spectra successfully matches the HR results
to percent level to $k_\mathrm{max}\approx16\,\hmpc$ at all redshifts,
a dramatic improvement over the LR predictions.
This is remarkable considering the fact that the SR model fares equally well
from the linear to the deeply nonlinear regimes.
While the model can learn and compare on even smaller scales using a larger FFT grid
(and a larger grid for density field as input to the discriminator
as described in the Materials and Methods),
the choice of $k_\mathrm{max}\approx16\,\hmpc$ is justified by the fact that
the model has reached an accuracy level comparable to that of the N-body simulations.

\subsection*{Halo mass function}\label{sub:hmf}

\begin{SCfigure*}[\sidecaptionrelwidth][t]
  \centering
  \includegraphics[width=0.8\textwidth]{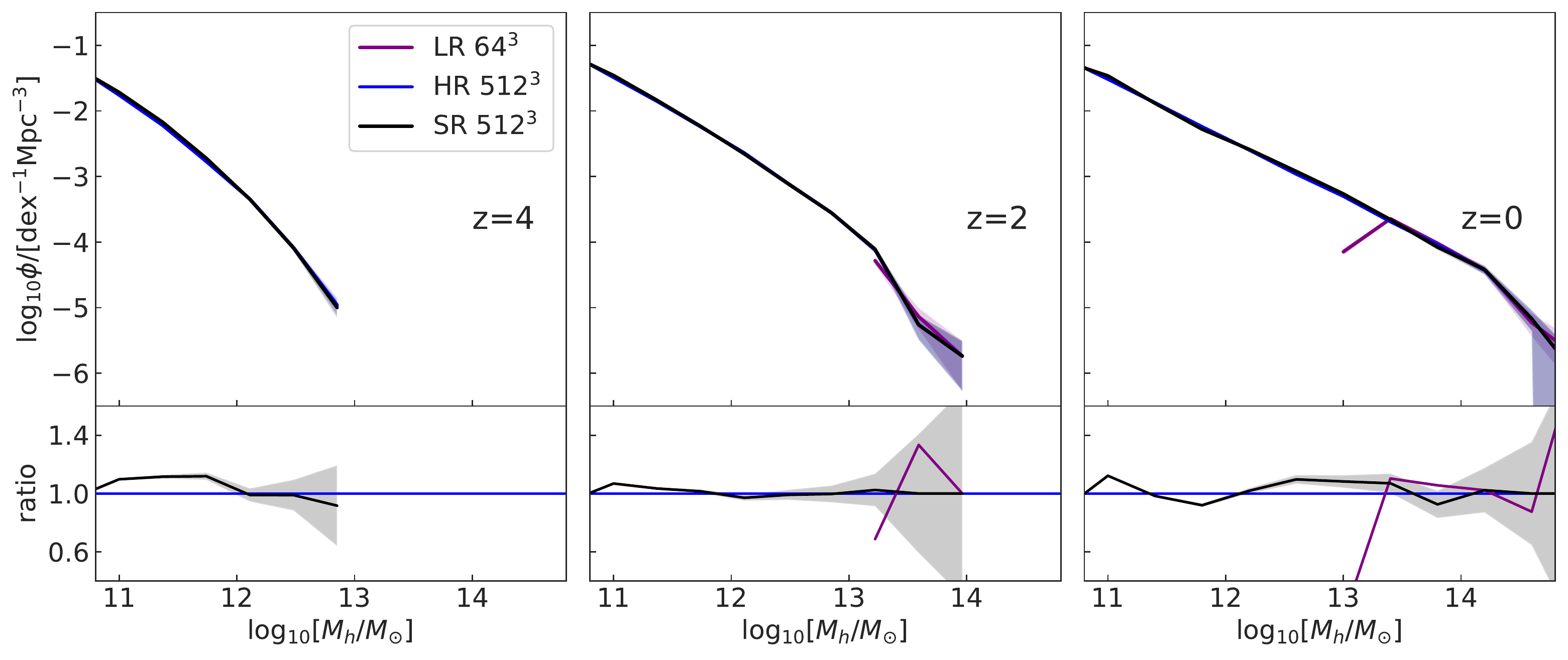}
  \caption{
  FOF Halo mass function comparison among LR (purple), HR (blue), and SR (black)
  test simulations, at $z=4, 2,$ and $0$.
  The LR simulations can only resolve the most massive halos
  ($M_{\mathrm h} \gtrsim 10^{13} \msun$).
  Our generated SR fields have the same mass resolution as the HR fields,
  resolving halos all the way down to $10^{11} \msun$.
  Their halo populations closely match the HR results,
  at the $10 \%$ level throughout the mass range.
  }\label{fig:hmf}
\end{SCfigure*}

Beyond the 2-point correlation test, i.e.\ the power spectrum,
we compare higher-order statistics of the
fields which quantify the non-Gaussianity arising in the nonlinear regime.
As the most nonlinear dark matter structures, halos host most of the
luminous extragalactic observables in the sky.
Therefore halo abundance is a natural choice when
considering non-Gaussian statistics.
We find the dark matter halos in the simulations using a
Friends-of-Friends (FOF) halo finder \citep{Davis85} with linking
length parameter $b = 0.2$.
The FOF algorithm links all pairs of particles within $b$ times their
mean separation, and collects each group of connected particles into one
halo.
We only keep halos with at least 32 particles.

In Fig.~\ref{fig:hmf} we compare the halo mass functions
with a $(100\,\hmpc)^3$ test set simulation, at $z=4, 2,$ and $0$.
The mass function is defined as
$\phi\equiv\d\,n / \d\log_{10}\!M_\mathrm{h}$,
where $n$ is the comoving number density of halos
above threshold mass $M_\mathrm{h}$.
Poisson errors on the halo abundance are shown.
These are conservative bounds because there are only small-scale
differences between the HR and SR
realizations  (see Fig.~\ref{fig:Realizations}) due to their conditioning on the same LR field.
Due to their large particle mass and low force resolution,
the LR simulations can only resolve the most massive halos above $10^{13} \msun$,
forming no halos at $z=4$.
The LR halo abundances are also underpredicted near the mass cut,
with deviation significantly larger than the Poisson error estimates.

Using our GAN model, the SR field generated from the LR input has the same mass resolution as the HR field, resolving halos over the whole mass range down to
$10^{11} \msun$.
The generated SR fields predict halo populations accurately compared with
the HR results, to $10 \%$ level through out the mass range.

\section*{Application to a large volume}\label{sec:1GPC}

\begin{figure*}
  \centering
  \includegraphics[width=1.0\textwidth]{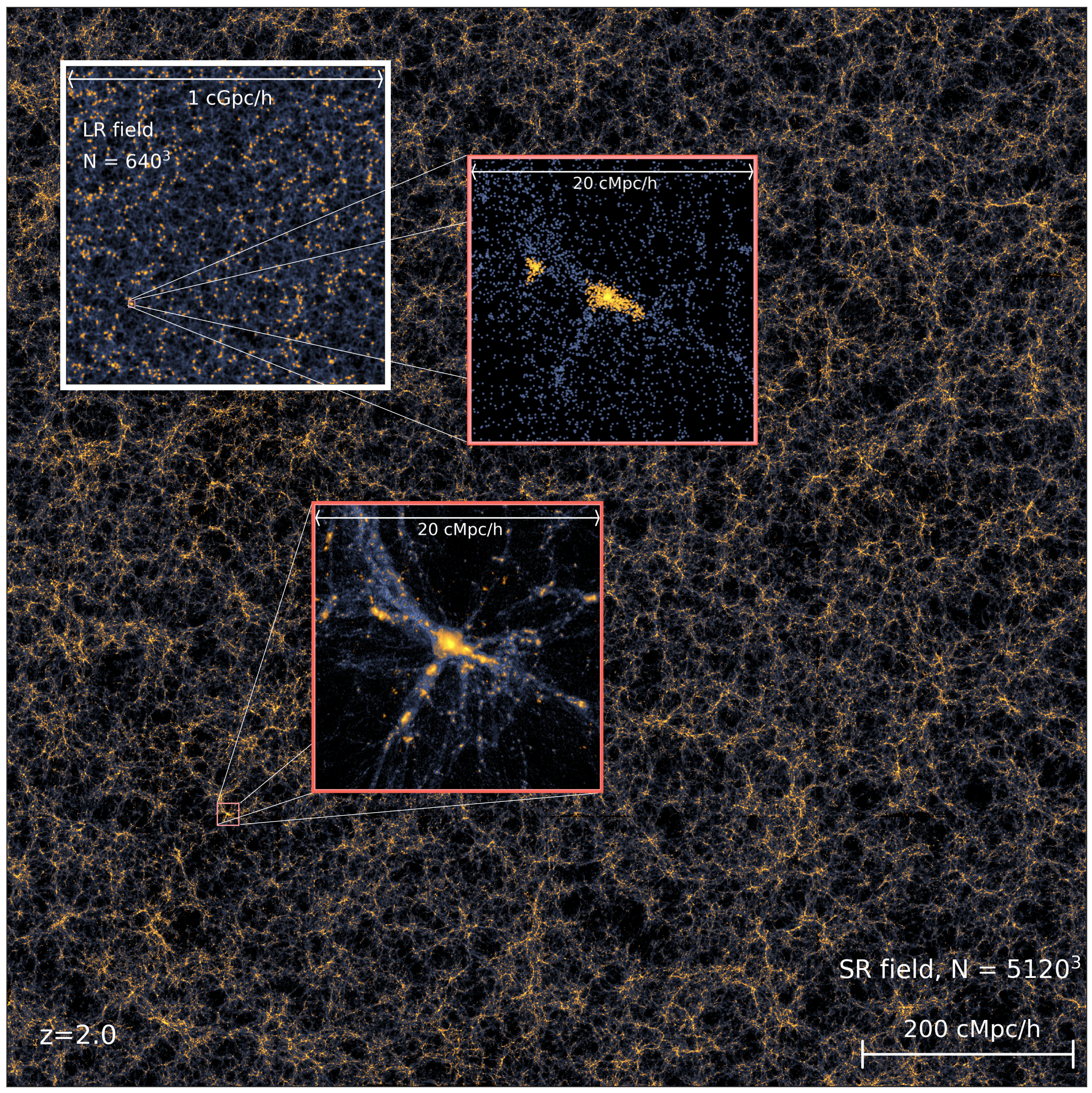}
  \caption{Illustration of the $(1 \hgpc)^3$ volume SR with $5120^3$ particles generated from the input $640^3$ LR field using our GAN model.
  Colors are the same as in Fig.~\ref{fig:slices}.
  The large panel shows a slice through the full box, 1 $\hgpc$ in length and 20 $\hmpc$ in thickness.
  The top left inset in the white box shows the same field from the LR simulation.
  In the two red inset panels, we show a $(20\,\hmpc)^3$ zoom-in region around a massive halo in both the SR and LR field.
  That central halo has a mass of $2 \times 10^{14} \msun$, larger than the most massive halo in the $(100\,\hmpc)^3$ volume training set.
  The generating process only takes about 16 hours on 1 GPU.
  }\label{fig:SR-1Gpc}
\end{figure*}

\begin{figure}[t]
  \centering
  \includegraphics[width=0.7\columnwidth]{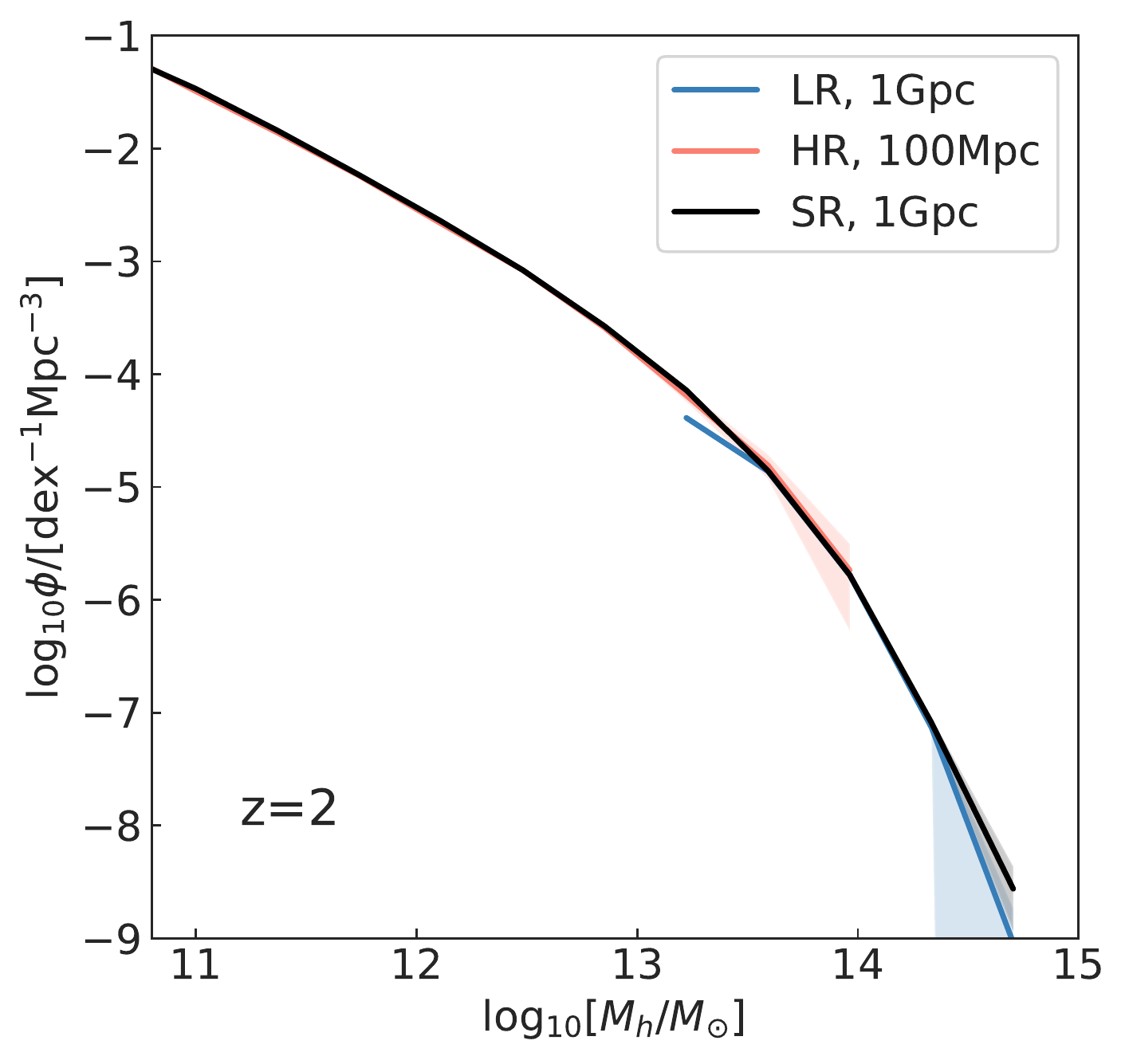}
  \caption{Halo mass function comparison of the $(1\,\hgpc)^3$ LR (blue),
  the $(1\,\hgpc)^3$ SR (black and generated from the former),
  and a $(100\,\hmpc)^3$ HR (orange) simulations at $z=2$.
  Limited by mass resolution, the LR run only resolves the most massive halos with $M_{\mathrm{h}} \gtrsim 10^{13} \msun$.
  Also, limited by volume, the HR run lacks halos above $M_{\mathrm{h}} \gtrsim 10^{14} \msun$.
  Our SR realization is able to match both the resolved LR and HR abundances
  over the whole mass range.
  We emphasize that our model generalizes surprisingly well to successfully predict
  the mass function of the massive halos that it has not been trained on,
  for over 3 orders of magnitudes in abundance,
  demonstrating its extrapolating power in modeling structure formations.
  }\label{fig:hmf-1Gpc}
\end{figure}

As described in the Materials and Methods, we train our GAN model on fields
cropped from $(100\,\hmpc)^3$ volume simulations.
In the previous section, our trained model has proven to work well on a
new SR realization of the same volume.
Moreover, translational symmetry has been carefully preserved by our cropping and padding schemes. This
implies that our method can be applied to any cosmological volumes.
The additional long-wavelength modes in the large volume lead to rare massive structures that the model has not encountered in the training set.
We apply our generative model to a $(1\,\hgpc)^3$ LR simulation with $640^3$ particles,
a volume 1000 times larger than the training set simulations,
and obtain an SR realization with $5120^3$ particles.
The generating process takes only about 16 hours on 1 GPU, including time for I/O.
This consumes significantly less computing resources than PDE-based N-body solvers.

Fig.~\ref{fig:SR-1Gpc} illustrates the generated $(1\,\hgpc)^3$ SR result at $z=2$,
teeming with finer details throughout the whole volume compared to the same LR field,
We zoom in to inspect a same halo in the SR and LR fields.
It has a mass of $2 \times 10^{14} \msun$,
larger than the most massive halo in the $(100\,\hmpc)^3$ training sets.
Remarkably, even though our GAN model has not been trained upon such massive halos, it performs well, and generates them with reasonable morphology, apparent substructure and correct population density (see below).

Quantitatively, we identify FOF halos in both the LR and SR large volumes,
and compare their resultant halo mass functions in Fig.~\ref{fig:hmf-1Gpc}.
Again the LR halos can only be well resolved above $10^{13} \msun$,
while the SR halo abundance agrees remarkably well with both the $(1\,\hgpc)^3$ LR
and the previous $(100 \,\hmpc)^3$ HR mass functions.
The SR prediction even has the right abundance for the massive halos above $10^{14} \msun$,
which are too rare to be present in the training volume.
Our model generalizes and extrapolates surprisingly well to predict
structure formations on scales over which it has not been directly trained.

This exercise demonstrates the potential of our newly developed method
to tackle the fundamental challenge of constructing large volume cosmological simulations at otherwise unfeasibly high resolution, with a dramatically lower footprint in computing resources.
It opens up an avenue for constructing large volume cosmological
(eventually hydrodynamical) simulations and creating large ensembles
of associated mocks at scales comparable to the current and future surveys,
for maximal scientific return.

\section*{Discussion}\label{sec:disc}

We have shown how a GAN super-resolution architecture inspired by StyleGAN2
can be used to enhance cosmological simulations so that they reproduce the
appearance and statistics of much higher-resolution models.
Our approach has similarities to the work of \cite{ramanah20},
who showed how SR dark matter density fields can be generated.
Among the differences with this work are our use of particle displacements
as the inputs and outputs of our modeling.
By generating a field of SR particle displacements, we effectively create
a whole simulation, which can be analysed (for example by carrying out halo
finding) in the same manner as a full simulation run at the higher resolution.
This in contrast with methods such as \cite{ramanah20} or
\cite{2018ComAC...5....4R}, which generate a density field.

One can ask whether there is fundamental limit to the level of
SR enhancement that our approach is able to produce.
In principle, one should be able to generate simulations completely using
a GAN, without a low-resolution model for conditioning, e.g.,
\cite{2018ComAC...5....4R, list20}.
Logically, such an approach would need proportionally more training data to
produce satisfactory results, as it would be generating large scale
modes as well.
At the opposite end, \cite{ramanah20} have added super-resolution details
to their simulations by enhancing by a factor of 2 in length scale.
In this paper we have shown that our SR modeling works well to produce
small-scale structures a factor of 8 in length (and 512 in mass)
below the LR scale.
At these scales one can begin to find bound structures such as subhalos.

The main advantage in the use of SR simulations over full HR runs is the potentially huge reduction in the computational resources involved.
As an example, for the $(100\,\hmpc)^3$ volume simulations of our test set, it takes approximately 560 core hours to run a $512^{3}$
particle HR model to $z=2$, while the $64^{3}$ particle LR run takes only about 0.6 core hours, a factor of $\sim 1000$ faster.
We can achieve even more dramatic speed up when applying the SR model to a larger volume.
For a $(1\,\hgpc)^3$ cosmological volume as shown above, only 500 core hours are needed to run a $640^{3}$ particle LR simulation to $z=2$.
To run a $5120^{3}$ HR counterpart with an N-body code would be daunting, and require dedicated supercomputing resources.
Our trained GAN model, on the other hand, only takes 16 hours with 1 GPU
(including the I/O time) to generate a $5120^{3}$ SR field
for the $(1\,\hgpc)^3$ volume,
a tiny fraction of the cost of the HR counterpart.
An additional advantage, made apparent by our processing the simulations in
distinct chunks is that the SR enhancement can be applied where it is needed,
e.g., in the vicinity of a specific galaxy cluster or supermassive black hole.
The data storage required can therefore also be much less than for a full HR run.

The stochastic nature of the StyleGAN implementation is also an interesting
feature which can be exploited.
As we have seen in Fig.~\ref{fig:Realizations}, it is possible to sample
multiple ``realizations'' of the small-scale clustering, conditioning on the large
scale modes, by varying the input noise component.
This opens up the possibility to improve the statistical inference of
cosmology from the small-scale clustering of galaxies, by jointly
sampling the small-scale modes with the cosmological parameters.

To train and test our model, we have used simulations of the same cosmology,
and this simplifies the learning task.
It is expected that the LR to HR mapping should depend weakly on cosmology,
and introducing such dependence should make the model adaptable and more
accurate for a range of parameters spanned by the training set.
Furthermore, to fully capture the state of a simulation, our model can be used
to learn particle velocities as well.
We leave those cosmology dependence and velocity improvements for future work.

A central question for GANs (and DL in general) is how to determine if enough
training data is being used.
Different diagnostics can be employed, and the dependence of accuracy on
training set size can be evaluated.
We have seen that our SR simulations are at least able to reproduce the power spectrum and halo mass
function well with a very limited set of training data.
A related issue is the fact that in astrophysics we often
deal with rare objects (e.g., quasars or galaxy clusters), and there may
not even be a single example in the training data.
In the future, one should examine how best to assemble a training dataset that
sufficiently covers all cases that will be studied.
As an example, separate universe or constrained simulations, e.g.,
\cite{li2014, barr19, huang20}
could be used to enhance the number of high mass objects in the training data.

So far though, we have demonstrated that our method is capable of generating SR simulations
one thousand times larger than the training sets.
Our large SR simulation does successfully reproduce the halo mass function and
power spectrum over several orders of magnitude.
It also leads to the formation of visually reasonable large and rare galaxy clusters.
Future work will further evaluate the quality of generated rare objects using other measures.

Particularly in the way we have generated them (as full particle
position datasets), the SR simulations could be used for many purposes that would
have needed HR simulations.
These include mock galaxy catalogues for large-scale structure studies,
e.g., \cite{lin20}.
Our current SR models are dark matter only, but could be used for mocks in the
same manner as HR models with resolved halos.
As we discuss below, extensions of this work will mean that it will be possible
to include hydrodynamic and star formation effects, through training using HR
models which have these physics incorporated.
SR mock catalogues could then be made which are more complex, for example
including reionization \cite{satpathy20}.
This type of mock making would have some similarities with the ``painting'' of
galaxies onto dark matter simulations by \cite{agarwal18}, except that unlike
that paper our method is conditioned on the entire density distribution rather than a few halo
properties, and would also add SR structure to the model.

Straightforward extensions of this work can be imagined,
with different levels of sophistication.
Particle data can be used to generate super-resolution
three dimensional gas, star and dark matter particle distributions.
Because the prediction will be of SR particles, these would effectively
function as full simulations, as we have mentioned previously.
In order to achieve this, training will have to be carried out on full physics
models, e.g., \cite{Feng2015, marshall2019}.
It should also be possible to train on HR models run with different
hydrodynamics algorithms \citep[e.g. ENZO by][]{Enzo} than are used in the LR models
(as long as the initial conditions are identical).
In such SR calculations, we will still be adding SR details to simulations after the LR models have been run, as in this paper.

Another extension is to replace the LR scheme, currently a full N-body solver, with faster alternatives,
such as another DL model, e.g., D3M \cite{he2019learning},
or a pseudo N-body solver, e.g., FastPM \cite{2016MNRAS.463.2273F} or FlowPM \cite{flowpm}.
The computational cost of a SR realization can be further reduced.
We will also be able to compute the full sensitivity matrix of the LR-SR model from back-propagation, which in turn allows us to use the method in Hamilton Monte-Carlo or similar sampling methods \cite[e.g.,][]{2017JCAP...12..009S,2019MNRAS.490.4237L} for a joint study of the initial condition and the cosmology parameters.

Beyond this, an improvement will be to run AI algorithms ``on the fly'', alongside the hydrodynamic simulation codes.
This will allow coupling and feedback between the physical processes at super resolved scales and
the LR scales.
For the most accurate modeling, this will be important, for example star
formation driven winds or metal pollution from supernovae can spread far from a
galaxy.
This means that the SR structure on galaxy scales should ideally be able to
affect the properties of the LR simulation.
To achieve this, it may also become necessary to carry out training on the fly
as HR and LR simulations run together.
We leave development of these methods to future work.

The ongoing revolution in Artificial Intelligence has already changed many fields of science.
Applications of these techniques are becoming widespread in cosmology also. Super-resolution
enhancement, that we have explored here, has many attractive features: Conventionally
generated large-scale Fourier modes are fully linked
to rapidly generated small-scale structures. This will allow consistent full volume simulations of the Universe
to be produced, covering an unprecedented dynamic range.

\matmethods{

\subsection*{A Survey of Models}\label{sub:models}

A variety of deep learning models have been applied to tackle
SR tasks, ranging from early convolutional neural
networks (CNN) plus simple loss function based supervised approaches
\citep[e.g., SRCNN;][]{dong2015}, to more recent unsupervised
methods such as Generative Adversarial Networks (GAN) \citep[e.g.,
SRGAN;][]{ledig2017photo}.

To generate SR outputs, both types of models use fully
convolutional networks, which apply multiple convolution layers with
learnable kernels of finite size.
The difference between the two lies in their loss functions.
By applying a simple loss function, e.g., the $\ell_2$ or $\ell_1$ norm,
on the difference between the HR output and target, the
first approach is a fully supervised learning task, and therefore easy to
train.
However the drawback is that those simple loss functions typically lead to
blurry output images, due to the fact that the target almost always
contains information not present in the input, i.e.\ high frequency
features in Fourier space.

As a remedy, more complex loss functions have been used to match the
output to the target on high-level features
\citep{johnson2016perceptual}.
Known as perceptual loss or content loss, it feeds the output and target
separately to another neural network (pre-trained on some other image
data set), and compares their feature maps at intermediate layers, still by
a simple loss function.
Because those intermediate results contains higher-level features
compared to the raw output or target, this method is able to generate
less blurry images.
However, such models are still deterministic and generate only one output for each
fixed input, whereas in principle each input could map to infinitely
many outputs due to the variability in the high-frequency modes.

The output image quality can be further improved by unsupervised
learning methods, such as GAN \citep{ledig2017photo}.
A GAN \citep{goodfellow2014generative} is a class of DL system in which
two NNs contest with each other in a game: the generative network
generates candidates while the discriminative network evaluates them.
Given a training set, this technique learns to generate new data with
the same characteristics as the training set.
GANs can be used to generate entirely new data from initially random
inputs, or in the case of SR, HR images from LR ones.
When the discriminator receives, in addition to SR or HR, the LR data as
input, it is able to identify true and generated
data conditioning on its large-scale knowledge.
This technique is known as the conditional GAN (cGAN)
\citep{mirza2014conditional}.

To solve the one-to-many problem in the input-to-target mapping, a model
needs more than the LR input to generate non-deterministic
output.
Different ways of introducing random factors have been attempted.
For example, the image-to-image translation model pix2pix
\citep{isola2017image} uses test-time dropout to add uncertainty to its
output.
More successfully, the image-generation models StyleGAN and StyleGAN2
\citep{karras2019style,karras2019analyzing} have achieved
state-of-the-art results on generating human
faces\footnote{\url{https://thispersondoesnotexist.com/}}.
It achieves a non-deterministic mapping by
simply adding noise after every convolution layer.
In our architecture, we use the same noise mechanism as in
StyleGAN2 to add stochasticity to the output.

In designing our neural network model, we are inspired mostly by
StyleGAN2.
Even though this model is originally designed for image-generation
tasks, it does so by successively upsampling some initial (constant) LR input,
a process extremely similar to our SR task.
We have also been influenced by the SR model SRGAN and the
image-to-image translation model pix2pix.

\begin{figure}[tb]
  \centering
  \parbox{\columnwidth}{
    \parbox[b]{0.62\columnwidth}{
      \subcaptionbox{generator}{
        \includegraphics[width=\hsize]{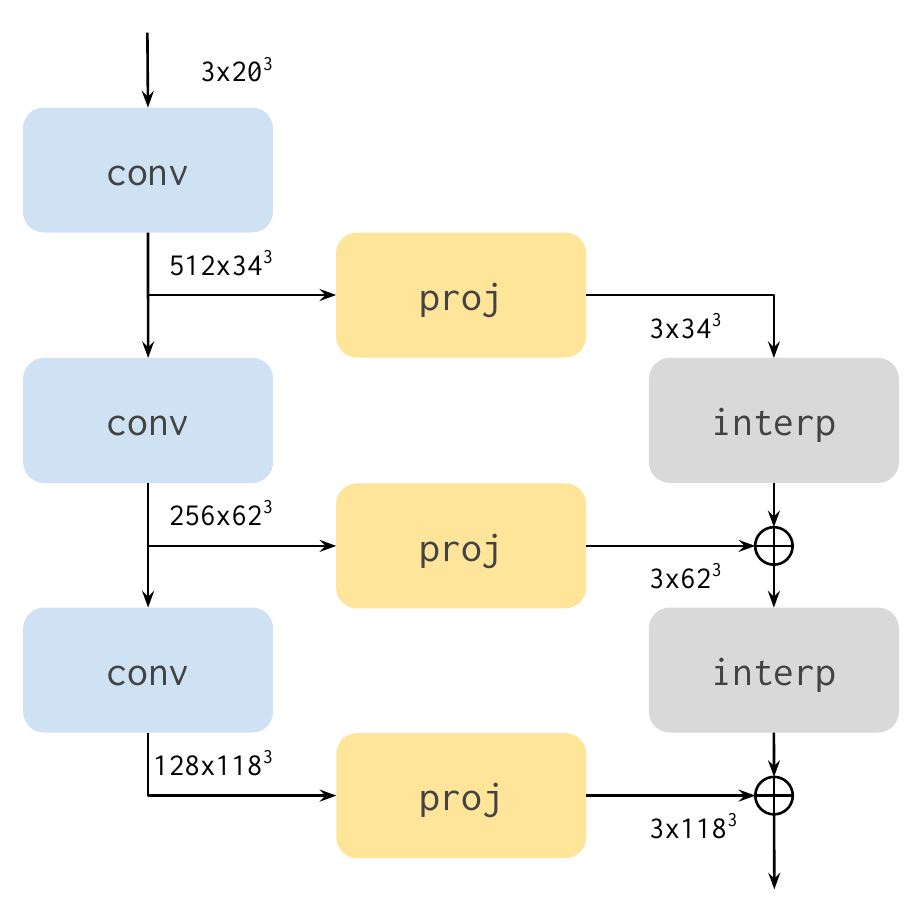}
      }
    }
    \hskip1em
    \parbox[b]{0.3\columnwidth}{
      \subcaptionbox{convolution block}{
        \includegraphics[width=\hsize]{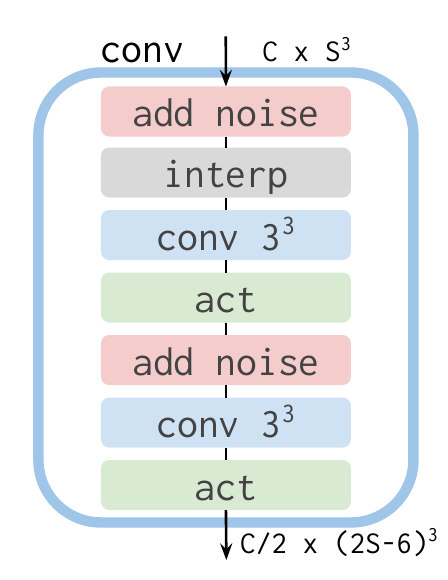}
      }
      \subcaptionbox{projection block}{
        \includegraphics[width=\hsize]{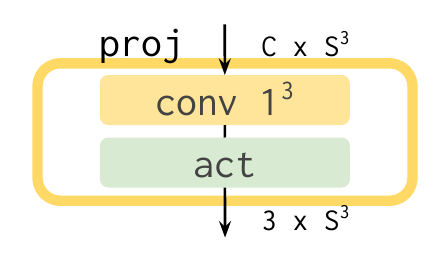}
      }
    }
  }
  \caption{Generator network architecture, inspired by StyleGAN2.
    The whole network structure is shown in (a), with components enlarged
    in (b) and (c).
    The colored plates are different operations, connected by lines or
    arrows from the input to the output.
    The sizes (channel number $\times$ spatial size) of the input,
    intermediate, and output tensors are next to the arrows.
    The generator takes the shape of a ladder, where each rung upsamples
    the data by $2\times$.
    The left rail consists of consecutive convolution blocks (``conv''
    in blue plates) operating in the high-dimensional latent space, and
    is projected (``proj'' in yellow plates) at every step to the
    low-dimensional output space on the right rail.
    The projected results are then upsampled by linear interpolation
    (``interp'' in gray plates), before being summed into the output.
    A key ingredient is the addition of noise (on red plates), that add
    stochasticity absent from the input at each level of resolution.
    The added noises are then transformed into high-frequency features by
    the subsequent convolutions and activations.
    The kernel sizes of the convolutions are labeled in their plates
    (that distinguish them from the ``conv'' block).
    Note that with a kernel size 1, ``conv $1^3$'' is simply an affine
    transformation along the channel dimensions, thus a convolution only
    in the technical sense.
    All activation functions (``act'' in green plates) are Leaky ReLU
    with slope $0.2$ for negative values.
    All ``conv'' blocks have the same structure as shown in (b) except
    the first one, which starts with an additional $1^3$ convolution and
    an activation.
  }\label{fig:generator}
\end{figure}

\begin{figure}[tb]
  \centering
  \parbox{\columnwidth}{
    \parbox[b]{0.23\columnwidth}{
      \subcaptionbox{discriminator}{
        \includegraphics[width=\hsize]{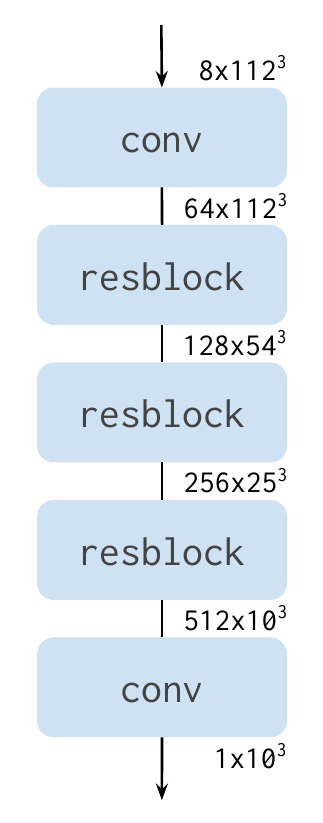}
      }
    }
    \hskip3em
    \parbox[b]{0.54\columnwidth}{
      \subcaptionbox{residual block}{
        \includegraphics[width=\hsize]{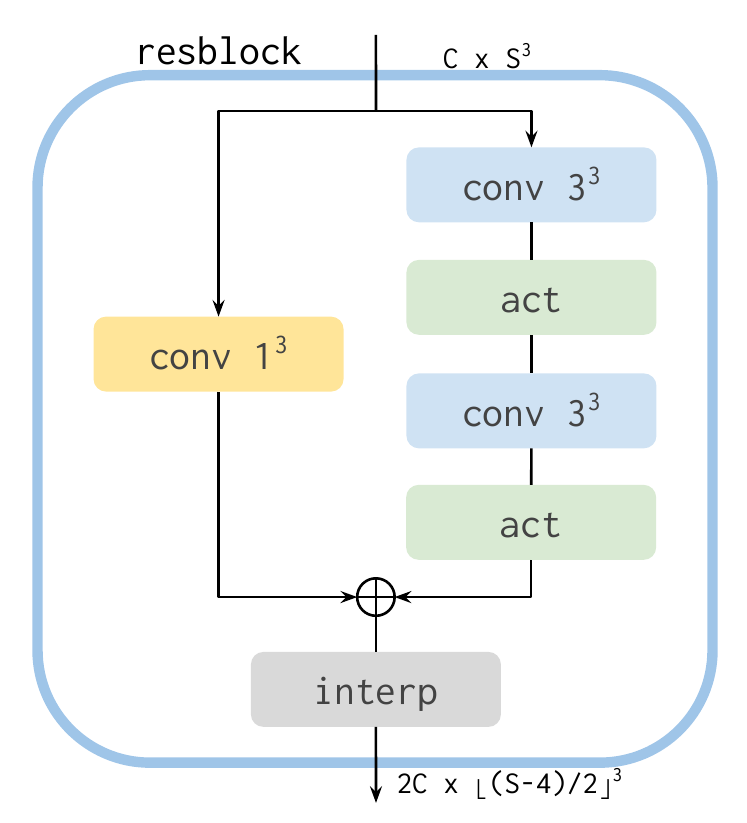}
      }
    }
  }
  \caption{Discriminator (critic) network architecture, inspired by the StyleGAN2.
    The whole network structure is shown in (a), with the residual block
    enlarged in (b).
    The residual block consists of two branches: the so-called ``skip''
    branch is only a $1^3$ convolution, and on top of that the other
    branch has real convolutions and activations to learn
    the ``residuals''.
    It then downsamples by $2\times$ the sum of the two branches with
    linear interpolation.
    Aside from the residual blocks, the first ``conv'' block include a
    $1^3$ convolution followed by an activation, and the last ``conv''
    block has a $1^3$ convolution to double the channel, an activation,
    and a $1^3$ convolution to reduce the channel to 1.
    See Fig.~\ref{fig:generator} for other details.
  }\label{fig:discriminator}
\end{figure}

\subsection*{Physical considerations}\label{sub:phys}

We perform the SR simulation task in the Lagrangian description.
Most often the particles are originally located on a uniform grid, so we
can structure a displacement field from the N-body particles as a 3D
image with 3 channels.
Each channel corresponds to one component of the displacement vector,
and the value in each voxel provides the displacement of the particle
originally from that voxel.

The deep learning model takes the LR particle displacements
as the input, and outputs a possible realization of their
HR counterparts.
Therefore, the outcome can be viewed as a higher-resolution simulation
with more particles and higher mass resolution.
Thus we name them the ``super-resolution simulations''.

Limited by the size of GPU memory and the fact that 3D data
consume more memory than lower dimensional tasks, we cannot feed the
whole simulations into the GPU during training and testing.
To overcome this, we crop each simulation into smaller spatial chunks.
While doing so it is desirable to preserve the translational symmetry
that arises naturally from the fully convolutional networks.
We explain the procedure below in more detail.

In addition to translational symmetry, the equations of motion of an
N-body system are also invariant under rotation.
Under the periodic boundary conditions imposed in numerical cosmological simulations,
the full 3D rotation group $\mathrm{SO}(3)$ reduces to a finite group of
48 elements, known as the octahedral group $\mathrm{O}_\mathrm{h}$.
Each group element can be decomposed into a permutation of the
simulation box axes and reflections along those axes.
During training, we feed the neural networks with input and output pairs
that are randomly transformed by these group elements.
Such data augmentation greatly enlarges the training data set, and
better enforces the said symmetry in the trained model.

With many more particles in the HR simulations, they can
contain small-scale information that is not present in their
LR counterparts.
To generate SR simulations that statistically match the
HR targets, a generator neural network requires extra
stochasticity in addition to that present in the LR input.
Also, this additional stochasticity needs to be transformed in a way such
that it results in the right correlations on different scales.
To this end, we add white noise to the intermediate feature maps, with
learnable amplitudes, throughout all stages of the neural network.
Noise injected at early stages is later upsampled, and can then
introduce correlations across multiple voxels.

\subsection*{Network architecture and loss function}\label{sub:net}

The design of our neural networks is mostly inspired by StyleGAN2
\citep{karras2019analyzing}.
The generator is illustrated in Fig.~\ref{fig:generator}.
The ladder-like structure of the generator upsamples the resolution by a factor of two
at each of its rungs.
See the figure captions for more architectural details.

With the LR simulations $l$ as inputs, our generator $G$
produces SR outputs $G(l)$ to mimic HR target
simulations $h$.
To train the generator, one simple option is to minimize the
voxel-by-voxel mean squared error (related to the $\ell_2$ norm) between $G(l)$
and $h$
\begin{equation}
  L_2 = \mathrm{E}_{l,h} \bigl[ \| G(l) - h \|_2 \bigr],
  \label{L2}
\end{equation}
where E$_{l,h}$ is the expected value while sampling all LR and HR
simulation pairs.
With the $L_2$ loss, the generator model is easier and faster to train,
and we can produce output tightly correlated with the HR
target on large scales.
However, below the input resolution scale, $L_2$ leads to blurry results
that lack high-frequency features, as explained above when introducing different DL models.

This problem can be addressed by replacing the simple loss function with
a discriminator network, which provides higher-level feedback to the
generator on its performance.
As shown in Fig.~\ref{fig:discriminator}, our discriminator utilizes
residual connections \citep{he2016deep} that perform well on
classification tasks.
See the figure captions for more architectural details.
Given either a generated SR or an original HR sample as input,
the discriminator $D$ learns to assign a score for it being real.
In the original GAN, this score is a probability to be compared to
the known values (0 or 1) using the binary cross entropy.

Wasserstein GAN \citep[WGAN][]{wgan} introduced a different score which
can be used to evaluate the wasserstein distance, the distance between
the distributions of real and fake images by optimal transport.
The discriminator (usually referred to as the critic in WGAN)
is subject to constraint of it being Lipschitz continuous with a
Lipschitz constant of 1, i.e.\ $|D(x_1) - D(x_2)| \leq |x_1 - x_2|$.
Compared to the vanilla GAN, WGAN is empirically superior for it requires
less tuning for training stability, and has a loss function that
converges as the generated image quality improves.
However it requires more computation per batch to maintain the Lipschitz
constraint.
This is most often achieved by adding a gradient penalty regularization
term to the WGAN loss function \citep[WGAN-gp;][]{wgan_gp}.
We train our networks using the WGAN-gp method, and only penalize the
critic gradient every 16 batches for training efficiency.

Because the HR and LR images must correlate on
large scales, and high-frequency features should depend on
the low-frequency context, we can make the discriminator
more powerful by giving it the LR images as additional input,
making our model a cGAN.
To achieve this we concatenate the upsampled LR images (by linear
interpolation) to both the SR output and HR target, respectively, as inputs
to the discriminator.
In addition, we also concatenate the density fields,
computed with a differentiable cloud-in-cell operation,
from all three displacement fields.
This allows the discriminator to see directly the Eulerian pictures
of formed structures, thus greatly enhancing its capability.
We find this addition crucial to generating visually sharp images
and accurate predictions of the small-scale power spectra.

The final adversarial loss function we use is
\begin{multline}
  L_\mathrm{WGAN-gp} = \mathrm{E}_{l} [D(l, G(l))]
  - \mathrm{E}_{l,h} [D(l, h)] \\
  + \lambda \; \mathrm{E}_{l, h} \bigl[\bigl( \|\nabla_i D(l, i)\|_2 - 1 \bigr)^2\bigr],
  %
  \label{L_GAN}
\end{multline}
where the first line gives the Wasserstein distance and the last term is
the gradient penalty.
$i$ is a random sample drawn uniformly from the line segment between
pairs of real ($h$) and fake ($G(l)$) samples.
We refer the readers to \cite{wgan_gp} for more details on WGAN-gp.
During training, we update the discriminator to minimize
$L_\mathrm{WGAN-gp}$, involving all three terms, and update the
generator to maximize it when only the first term takes effect.
%

In both the generator and the discriminator, we use convolution layers
with no padding.
Recall that the inputs and outputs of our networks are cropped parts of
the bigger simulations, as limited by the size of GPU memory.
Convolutions with zero padding or other forms of non-periodic padding break
translational invariance.
Without padding in the convolution layers, the outputs of the
generator are smaller than a simple factor-of-8 scaling.
We compensate for this by adding extra padding to the inputs of the
generator.

\subsection*{N-body Simulation Dataset}\label{sub:sim}

To train and validate our SR model, we use N-body simulations which only
contains dark matter interacting via gravity.
The dynamics of dark matter are evolved using
\texttt{MP-GADGET}~\citep{Feng2015, Feng2016}.
It is an N-body and hydrodynamics cosmological simulation code
optimized to run on the most massively parallel high performance
computer systems.
MP-Gadget was used to run the BlueTides simulation \citep{Feng2016}
on BlueWaters, the only cosmological hydrodynamic code that has
carried out full machine runs, scaling to 648,000 cores and producing
more than 6 petabytes of usable data.
In all simulations the gravitational force is solved with a split
Tree-PM approach, where the long-range forces are computed from a
particle-mesh method and the short-range forces are obtained with a
hierarchical octree algorithm.

We use $1/30$ of the mean spatial separation of the dark matter particles
as the gravitational softening length.
The simulations have the WMAP9 cosmology with matter density $\Omega _{\rm m} =
0.2814$, dark energy density $\Omega _{\Lambda} = 0.7186$, baryon
density $\Omega _{\rm b} = 0.0464$, power spectrum normalization $\sigma
_{8} = 0.82$, spectral index $n_{s} = 0.971$, and
Hubble parameter $h = 0.697$.
We train our model separately on simulation snapshots,
at redshift $z=4, 2,$ and $0$, of different levels of nonlinearity.
We expect that it is harder for the model to learn to form nonlinear
structures than linear ones, and these redshifts allow us to test
the performance degradation due to nonlinearity.

For training and testing we run, respectively, 16 and 1 LR-HR pairs of dark matter only simulations with box size of $(100\,\hmpc)^3$, and $64^3$ and $512^3$ particles for LR and HR respectively.
The mass resolution is $m_{\mathrm{DM}} = 2.98 \times 10^{11} \msun/h$ for LR and $m_{\mathrm{DM}} = 5.8 \times 10^{8} \msun/h$ for HR.
So our SR task is to enhance the spatial resolution by $8\times$ and mass resolution by $512\times$.
We also run a $(1\,\hgpc)^3$ LR simulation, of otherwise the same
configuration as the smaller LR runs, to test deploying our model
to a larger volume.

\subsection*{Training and Testing}\label{sub:train}

Our GAN model is trained upon the displacement field in Lagrangian space.
We first pre-process the training data by converting the particle position to their displacement vector with respect to the initial grid.
Then for each mini-batch, we crop a $14^3$ grid from the LR displacement field as input, pad 3 cells on each side to compensate for the loss of voxels during the generator convolution layers.
This transforms into an SR output of size $118^3$ through the generator
(see sizes annotated in Fig.~\ref{fig:generator}),
out of which we crop the inner part to match the corresponding $(8\times14)^3 = 112^3$ grid from the HR target displacement field.
Therefore, all the mini-batches for training are about $(22\,\hmpc)^3$ in size.
We then concatenate the tri-linear interpolations ($8\times$ upsampling
per dimension) of the LR inputs to the SR outputs or the HR targets.
The results are 6 channel images to be taken by the discriminator
and transformed in a similar way as shown in Fig.~\ref{fig:discriminator}. Data augmentation is applied for each mini-batches.

We train our neural network model using the Adam optimizer with learning
rate $1 \times 10^{-5}$ and exponential decay rates $\beta_1 = 0$ and
$\beta_2 = 0.99$.
We start with 5 epochs of supervised training with the simple loss
function given in Eq.~\ref{L2}.
This allows the generator to quickly learn to generate SR outputs that are
consistent with LR inputs above the input resolution.
We then proceed to the adversarial training and update the generator and
the discriminator alternately, for 150 epochs.

For the test set, we use a new pair of $64^3$ LR and $512^3$ HR simulation with same volume of $(100\,\hmpc)^3$.
The test set shares the same cosmological parameters as the training sets, but is a new realization with different initial conditions.
Therefore our comparison is free of overfitting to the training data set.
Remarkably, we are able to deploy our model to LR input of $1000\times$ bigger volume
and demonstrate the scaling ability of our approach.

The procedure of generating the SR simulation from the LR input goes as follows.
We first preprocess the LR input by converting the particle position to the displacement field of shape $3 \times 64^3$.
Then we use our trained GAN model to generate the SR displacement field with shape $3 \times 512^3$, and obtain the particle positions by moving them from their original positions on a lattice by these displacement vectors.
The generation process from LR to SR is done in chunks due to the limit of GPU memory. We crop the LR input into pieces, generate their corresponding SR field, and stitch the output patches together to exactly match the shape of the target.
The generated full SR field is periodically continuous through this
procedure thanks to the translational symmetry preserved by our GAN model.

We have implemented our model and training in \texttt{map2map}
(\url{https://github.com/eelregit/map2map}),
a general neural network framework to transform field data.
}

\showmatmethods{}

\acknow{The Flatiron Institute is supported by the Simons Foundation.
We acknowledge the Texas Advanced Computing Center (TACC) at The
University of Texas at Austin for providing HPC resources that have
contributed to the research results reported within this paper.
TDM acknowledges funding from NSF ACI-1614853, NSF AST-1616168, NASA ATP 19-ATP19-0084 and 80NSSC20K0519.
TDM and RACC also acknowledge funding from NASA ATP 80NSSC18K101, and NASA ATP NNX17AK56G, and RACC, NSF AST-1909193.
SB was supported by NSF grant AST-1817256.
This work was also supported by the NSF AI Institute: Physics of the Future, NSF PHY-2020295.
}
\showacknow{}

\bibliography{main}

\end{document}